\documentclass[twocolumn,showkeys,prd,nofootinbib,floatfix]{revtex4-1}

\usepackage{amsfonts,amsmath,amssymb} \usepackage{graphicx,graphics,color}

\def\mnras{Mon. Not. Roy. Astron. Soc.} 
\def\mnras{MNRAS} \def\apj{ApJ} \def\apjl{ApJ} 
  \def\aap{A\&A} 
  
  \def\prd{Phys. Rev. D}
  
 \def\nat{Nature} 
\def\pasj{PASJ}

\begin{document}

\title{Strong lensing by fermionic dark matter in galaxies}

\author{L.~Gabriel G\'omez,$^{1,2,3}$ C. R. Arg{\"u}elles,$^{3,4}$ Volker
Perlick,$^{5}$ J.~A.~Rueda,$^{1,3,6}$ R.~Ruffini$^{1,3,6}$}

\affiliation{$^1$Dipartimento di Fisica and ICRA, Sapienza Universit\`a di
Roma, P.le Aldo Moro 5, I--00185 Rome, Italy} \affiliation{$^2$University
of Nice-Sophia Antipolis, 28 Av. de Valrose, 06103 Nice Cedex 2, France}
\affiliation{$^3$ICRANet, Piazza della Repubblica 10, I--65122 Pescara,
Italy} \affiliation{$^4$Grupo de Astrofisica, Relatividad y Cosmolog\'{i}a,
Facultad de Ciencias Astronomicas y Geofisicas\\ Universidad Nacional de La
Plata and CONICET, Paseo del Bosque S/N 1900 La Plata, Pcia. de Buenos
Aires, Argentina} \affiliation{$^5$ZARM, University of Bremen, 28359
Bremen, Germany} \affiliation{$^6$ICRANet-Rio, CBPF, Rua Dr. Xavier Sigaud
150, Rio de Janeiro, RJ, 22290--180, Brazil}

\date{\today}

\begin{abstract} It has been shown that a self-gravitating system of
massive keV fermions in thermodynamic equilibrium correctly describes the
dark matter (DM) distribution in galactic halos (from dwarf to spiral and
elliptical galaxies) and that, at the same time, it predicts a denser
quantum core towards the center of the configuration. Such a quantum core,
for a fermion mass in the range of $50$~keV~$\lesssim m c^2 \lesssim
345$~keV, can be an alternative interpretation of the central compact
object in Sgr A*, traditionally assumed to be a black hole (BH). We present
in this work the gravitational lensing properties of this novel DM
configuration in nearby Milky Way-like spiral galaxies. We
describe the lensing effects of the pure DM component both on halo scales,
where we compare them to the effects of the Navarro-Frenk-White and the
Non-Singular Isothermal Sphere DM models, and near the galaxy center, where
we compare them with the effects of a Schwarzschild BH. For the particle
mass leading to the most compact DM core, $m c^2\approx 10^{2}$~keV, we draw
the following conclusions. At distances $r\gtrsim 20$~pc from the center of
the lens the effect of the central object on the lensing properties is
negligible. However, we show that measurements of the deflection angle
produced by the DM distribution in the outer region at a few kpc, together
with rotation curve data, could help to discriminate between different DM
models. In the inner regions $10^{-6}$~pc~$\lesssim r\lesssim 20$~pc, the
lensing effects of a DM quantum core alternative to the BH scenario, becomes a theme of an analysis of unprecedented precision
which is challenging for current technological developments. We show that
at distances $\sim 10^{-4}$~pc strong lensing effects,
such as multiple images and Einstein rings, may occur. Large differences
in the deflection angle produced by a DM central core and a central BH
appear at distances $r\lesssim 10^{-6}$~pc; in this regime
the weak-field formalism is no longer applicable and the exact
general-relativistic formula  has to be used for the deflection angle which
may become bigger than $2 \pi$. An important difference in comparison to
BHs is in the fact that quantum DM cores do not show a photon sphere; this
implies that they do not cast a shadow (if they are transparent). Similar
conclusions apply to the other DM distributions for other fermion masses in
the above specified range and for other galaxy types. \end{abstract}

\pacs{Valid PACS appear here}
\maketitle

\section{Introduction}
Most of the intriguing problems in particle physics and cosmology are
related to the nature of the dark matter (DM) that composes approximately
$80\%$ of matter in the Universe \cite{2016A&A...594A..13P}. In a
self-consistent particle DM model aimed to understand the quasi relaxed DM
halo structures, the underlying phase-space distribution and
self-gravitation establishes how the DM distributes in the galaxy. Namely,
the distribution of DM inside the galaxy (e.g. mass, density and pressure
profiles) can be obtained from the solution of the hydrostatic equilibrium
equations and the corresponding (phase-space dependent) equation of state
(see e.g.~\cite{2008gady.book.....B}).

Before a DM halo enters in the steady state we observe, and due to the
collisionless nature of the DM particles, specific relaxation mechanisms
such as violent relaxation take place in a few dynamical times, giving rise
to quasi-stationary states which can be described by (coarse grained)
phase-space distributions of the Fermi-Dirac type
\cite{1998MNRAS.296..569C,2004PhyA..332...89C}. Indeed, it has been
recently shown that a model of DM based on a self-gravitating system of
fermions in thermodynamic equilibrium accurately describes the distribution
of DM in galactic halos, when contrasted with observations \cite{2015MNRAS.451..622R,2015ARep...59..656S,2016arXiv160607040A}). In Ref.~\cite{2015MNRAS.451..622R}, it was shown that
such a self-gravitating system of fermionic DM shows a general DM density
distribution, hereafter the RAR profile, with a \emph{compact core} -
\emph{diluted halo} structure (see Sec.~\ref{sec:2} for details). More
recently, following the more complex and realistic statistical approach
accounting for escape of particles, in Ref.~\cite{2016arXiv160607040A} a cutoff in the fermion momentum distribution was introduced. Such a
momentum cutoff serves to account for the finite size of galaxies,
generalizing the previous RAR profile \cite{2015MNRAS.451..622R}. Both
Refs.~\cite{2015MNRAS.451..622R} and \cite{2016arXiv160607040A} have put
constraints on the mass of these fermions, hereafter called \emph{inos},
using known observational properties of galaxies such as the flatness of
the rotation curves, the mass and radius of galaxies, as well as
observationally-inferred correlations involving many different galaxy
types.

On the other hand, gravitational lensing (GL) has been widely used to
determine the distribution of DM in galaxies and galaxy clusters
\cite{2001ApJ...552..493S,1991ApJ...370....1M}. Hence, given a specific
density profile it is systematically possible to infer the GL properties
for any lens system or vice versa, i.e., if we know the lensing signal, we
can reconstruct the mass distribution of the lens under some assumptions of it. Moreover, in \cite{2011ApJ...729..127U} a Bayesian statistical method was presented that permit us to reconstruct a model independent mass profile without initial assumptions by combining measurements of magnification bias along with lens distortion. For instance, the gravitational lensing
properties given by the phenomenological Navarro-Frenk-White (NFW) profile,
commonly used to describe the cold dark matter (CDM) distribution of halos,
have been very well investigated (see, e.g.,
Ref.~\cite{2003MNRAS.340..105M}, and references therein). The same applies
to the non-singular isothermal sphere (NSIS) profile
\cite{1987ApJ...320..468H}. Interestingly, the lensing data are better
fitted by the latter kind (see e.g.~\cite{2001ApJ...561...46K}), which is
cored-like (i.e. with a shallower inner DM halo density profile in contrast
to the more cuspy NFW one), in a way similar to the RAR profile in that
galaxy region (see Fig. 3 in \cite{2015MNRAS.451..622R}).
%
Particular attention has been paid in the last decade to single galaxies
where strong lensing effects are commonly present. Surveys such as SWELLS
\cite{2011MNRAS.417.1621D,2012MNRAS.423.1073B} and DiskMass
\cite{2013A&A...557A.131M}, among others, have placed constraints on the
properties of spiral galaxies, revealing for instance the DM fraction
within 2.2 disk radii, the inner logarithmic slope of the DM halo profile,
as well as the stellar mass component (disk plus bulge). Including
kinematic analysis, it is also possible to break the \textit{disk-halo
degeneracy}  \cite{2011MNRAS.417.1621D} and to put a more stringent
constraint on the aforementioned properties (see also
\cite{1998ApJ...495..157K} for a theoretical study and
\cite{2015ApJ...801L..20C} for a summary of the DM properties). Likewise,
the slope of the average DM projected density profile in the innermost
regions of massive early type galaxies, has been constrained by using strong
gravitational lensing data along with stellar dynamics and stellar
population
\cite{2012ApJ...747L..15G,2012ApJ...752..163S,2014MNRAS.445..115T}.
Recently, the DM density profile has been constrained at a few kpc,
contributing significantly in the same way as the stellar component does
\cite{2015ApJ...799L..22H,2015MNRAS.449.2128K}, exhibiting that these types
of galaxies have a non-negligible amount of DM in their central regions.

Observations also indicate that most galaxies host a massive central
compact object, usually assumed to be a black hole (BH). Its presence
substantially affects lensing features such as the \emph{critical curves}
and the formation (suppression) of an additional (existing) faint central
image, as  predicted in \cite{2001MNRAS.323..301M} and already observed \cite{2004Natur.427..613W}.
Remarkably, such effects depend on the halo core radius and a critical
value of the mass of the BH \cite{2015ApJ...799L..22H,2001MNRAS.323..301M},
leading to a strong degeneracy between these parameters.

At this point we turn back to the discussion of DM models on galaxy scales
to recall an interesting feature of the RAR model
\cite{2016arXiv160607040A}: its DM central core, hereafter DMCC,  can be
compact enough to correctly describe the observational properties in the
Galactic center, which are usually associated with the existence of a
massive BH centered in Sgr A*. Namely, it can produce a gravitational
potential which suffices to explain the dynamics of the stars closest to
Sgr A*, the S-cluster stars (see Ref.~\cite{2016arXiv160607040A} for
details). Thus, the RAR profile could, in principle, explain the MW
properties from the center all the way to the halo.

It is thus natural to ask ourselves about the GL properties of nearby
lensing galaxies modeled within the RAR model. Without loss of generality,
we compute in this work the GL properties of DM halos for spiral type
galaxies such as the MW. In addition to the inclusion of the halo part,
which is slightly but appreciably distinguishable from other DM models,
we consider the lensing effects of the DM distribution
near the GC where a maximum deflection of light is predicted, in contrast
to standard models of DM such as NFW and NSIS. We also show that, at such
scales, the deflection angles are no longer small so that the exact
equations from general relativity must be used.

This work is organized as follows: In Sec.~\ref{sec:2} we describe the
general features of the novel model of DM fermions and compare them with
the ones obtained by the standard density profiles. We compute in
Sec.~\ref{sec:3} the gravitational lensing properties of our DM
distribution in Milky Way-like galaxies and compare them
in the halo with those of the NFW and NSIS profiles and in the core with
those of a Schwarzschild BH as they have been predicted for Sgr A*
\cite{2000PhRvD..62h4003V}. Finally we present a general discussion in
Sec.~\ref{sec:4} of the GL properties of the fermionic DM distribution in
galaxies.

\section{The density Profiles}\label{sec:2}
We first recall the widely used phenomenological DM density profile arising
within the $\Lambda$CDM cosmological paradigm, i.e. the NFW profile
\cite{1996ApJ...462..563N}
\begin{equation}
\rho(r)=\frac{\rho_{c}}{(r/r_{s})(1+r/r_{s})^{2}},\label{a1} \end{equation}
where $\rho_{c}$ is the characteristic density and $r_s$ is the scale
radius. This density profile exhibits a sharp cusp in the inner region
$\rho\varpropto r^{-1}$ while in the halo part the density scales as
$\rho\varpropto r^{-3}$.

Another often adopted DM density profile which also yields the asymptotic
flatness of the rotation curves is represented by the \emph{non-singular
isothermal} (NSIS) profile \cite{1999MNRAS.307..203S}:
\begin{equation} \rho(r)=\frac{\rho_{0}}{1+(r/r_{0})^{2}},\label{a2}
\end{equation}
where $\rho_{0}$ is the central density and $r_{0}$ is the core radius.

We turn now to the RAR profile
\cite{2015MNRAS.451..622R,2016arXiv160607040A} which describes the DM
distribution along the entire galaxy in a continuous way, i.e. from the halo
part to the GC and without spoiling the baryonic component which dominates
at intermediate scales (see \cite{2016arXiv160607040A} for details).

\begin{figure} \centering \includegraphics[width=\hsize,clip]{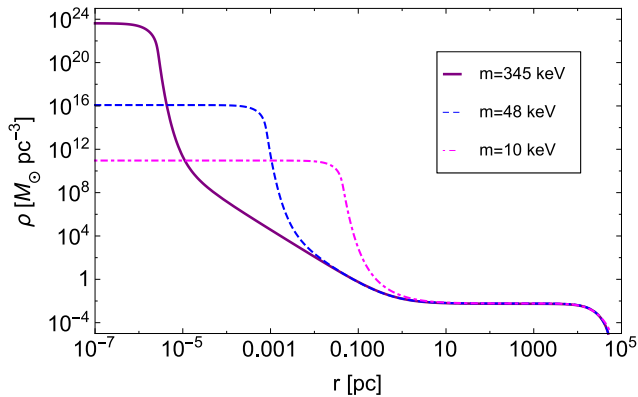}
\caption{Density profiles for MW-type galaxies for different particle mass
values that satisfy the boundary conditions: $M_{DM}(r=20 kpc)=9\times
10^{10}M_{\odot}$ (or equivalently $M_{DM}(r=40 kpc)=2\times
10^{11}M_{\odot}$), with a DMCC of mass $M_c=4.2\times 10^{6} M_{\odot}$.
The set of initial conditions: $\theta_{0} = 32.2; \beta_{0}=
2.7\times10^{-7}; W_{0}=58.5$, $\theta_{0}= 37.1; \beta_{0}=1.04\times
10^{-5}; W_{0} = 65,20$, $\theta_{0} = 45.8; \beta_{0}= 4.0\times10^{-3};
W_{0}=76.049$ for $m=10$~keV, $m=48$~keV and $m=345$~ keV, respectively,
have been here adopted from the original solutions computed in
\cite{2016arXiv160607040A}. ß««} \label{fig:inosdensity}
\end{figure}

Assuming a self-gravitating system of massive fermions (within the standard
Fermi-Dirac phase-space distribution) in thermodynamic equilibrium, the DM
density profile was computed in \cite{2015MNRAS.451..622R}. By imposing
fixed boundary conditions at the halo and including the fulfillment of the
rotation curve data, the parameters of the system have been constrained.
This procedure was applied for different types of galaxies from dwarfs to
big spirals exhibiting a universal \emph{compact core} - \emph{diluted
halo} density profile. An extended version of the RAR model was recently
presented \cite{2016arXiv160607040A}, by introducing a fermion energy
cutoff $\epsilon_{c}$ in the fermion distribution. This is also motivated
by the formal stationary solution (Fermi-Dirac-like) of the generalized
statistics which includes the effects of escape of particles and violent
relaxation \cite{2004PhyA..332...89C}. The new emerging density profile
serves to account for the finite galaxy sizes due to the more realistic
boundary conditions, while also opening the possibility to achieve a more
compact solution for the quantum core working as a good alternative to the
BH scenario in Sgr A* (see Fig.~\ref{fig:inosdensity}).

Motivated by these features between the DM profiles, we compute in the next
section both the GL properties of MW-type galaxies for the RAR model and
the GL effect of the DMCC in order to study the possibility of strong
lensing effects around the GC. We describe now the RAR profile following
Ref.~\cite{2016arXiv160607040A}.

The density $\rho$ and pressure $P$ for this system are given by
\begin{eqnarray} \rho&=&\frac{g}{h^{3}}m \int_{0}^{\epsilon_{c}} f_{c}(p)
\left(1+\frac{\epsilon(p)}{mc^{2}}\right)d^{3}p,\label{b4}\\
P&=&\frac{2}{3}\frac{g}{h^{3}} \int_{0}^{\epsilon_{c}} f_{c}(p)
\frac{1+\epsilon(p)/2mc^{2}}{1+\epsilon(p)/mc^{2}}d^{3}p,\label{b5}
\end{eqnarray}
where $g$ denotes as usual the particle spin degeneracy and $f_{c}(p)$ is
the phase space distribution function including an energy cutoff:
\begin{equation} f_{c}(p)= \left\{ \begin{array}{ll}
\frac{1-e^{(\epsilon-\epsilon_{c})/kT}}{e^{(\epsilon-\mu)/kT}+1}  &
\epsilon\leqslant \epsilon_{c}, \\ 0 &  \epsilon >\epsilon_{c}. \end{array}
\right. \end{equation}
Here $\epsilon=\sqrt{c^{2}p^{2}+m^{2}c^{4}}-mc^{2}$ is the particle kinetic
energy, $m$ is the particle mass, $\mu$ is the chemical potential (with the
particle rest mass subtracted off), $T$ is the temperature, $k$ is the
Boltzmann constant, and $h$ is the Planck constant.

Considering the spherically symmetric space-time described by the metric
\begin{equation}
ds^{2}=e^{\nu}dt^{2}-e^{\lambda}dr^{2}-r^{2}d\Theta^{2}-r^{2} \mathrm{sin}
^{2} \Theta \, d\phi^{2}, \end{equation}
along with the thermodynamic equilibrium conditions $e^{\nu/2}T=$~const.
and $e^{\nu/2}(\mu+mc^{2})=$~const. and the equation of state given by
Eqs.~(\ref{b4}) and (\ref{b5}), the  dimensionless Einstein equations are
finally obtained (see Ref.~\cite{2016arXiv160607040A}, for details),
\begin{eqnarray}
&&\frac{d\hat{M}}{d\hat{r}}=4\pi\hat{r}^{2}\hat{\rho},\label{b8} \\
&&\frac{d\theta}{d\hat{r}}=-\frac{[1-\beta_{0}(\theta-\theta_{0})]}{\beta_{0}}\frac{\hat{M}+4\pi
\hat{P}\hat{r}^{3}}{\hat{r}^{2}(1-2\hat{M}/\hat{r})},\label{b9} \\
&&\frac{d\nu}{d\hat{r}}=2\frac{\hat{M}+4\pi
\hat{P}\hat{r}^{3}}{\hat{r}^{2}(1-2\hat{M}/\hat{r})},\label{b10} \\
&&\beta(r)=\beta_{0}e^{\frac{\nu_{0}-\nu(r)}{ 2}},\label{b11} \\
&&W(r)=W_{0}+\theta(r)-\theta_{0},\label{b12} \end{eqnarray}
where we have introduced the temperature parameter $\beta=k T/mc^{2}$, the
degeneracy parameter $\theta=\mu/kT$ and the cutoff parameter
$W=\epsilon_{c}/kT$. In addition, we have introduced the
dimensionless quantities $\hat{r}=r/\chi$, $\hat{M}=GM/(c^{2}\chi)$,
$\hat{\rho}=G\chi^{2}\rho/c^{2}$ and $\hat{P}=G\chi^{2}P/c^{4}$ with
$\chi=2\pi^{3/2}(\hbar/mc)(m_{p}/m)$ and  $m_{p}$ being the Planck mass.
The subscript 0 in the Einstein equations denotes the initial condition
values. Thus, the fermionic  DM density profile was computed numerically
and characterized by a quantum central core of almost constant density, an
intermediate transition region followed by an extended plateau and a
Boltzmannian density tail $\rho\varpropto r^{-\alpha}$, with $\alpha>2$ due
to the cutoff condition (see Fig.~\ref{fig:inosdensity}). This latter
feature, together with the addition of the standard baryonic disk
component, leads to the observed flat rotation curves (see
\cite{2016arXiv160607040A} for the specific analysis of the MW). Based on
this property, a family of solutions can be obtained to guarantee the
fulfillment of both the boundary conditions of the halo part and rotation
curve data, with different central quantum cores of different compactness
presenting a clear dependence on the fermion mass (see
Fig.~\ref{fig:inosdensity} for the case of the three solutions of
interest).

The size of the degenerate quantum core depends on the particle mass as can
be seen in Fig.~\ref{fig:inosdensity} for the case of the MW.
As shown in \cite{2016arXiv160607040A}, for an ino mass in the range of
$48$ keV$\lesssim m \lesssim 345$~keV, the DM core has a size (and a mass)
appropriate to describe the orbit of the S-cluster stars around Sgr A*
\cite{2009ApJ...707L.114G}. Thus, the DM core represents a valid
alternative to the central BH hypothesis. In addition, for such fermion
masses the DM contribution properly reproduces the total rotation curve
data, without spoiling the baryonic components available above parsec scale
(see \cite{2016arXiv160607040A} for details and see
\cite{2013PASJ...65..118S} for the latest data available in the MW).

\begin{figure} \centering \includegraphics[width=\hsize,clip]{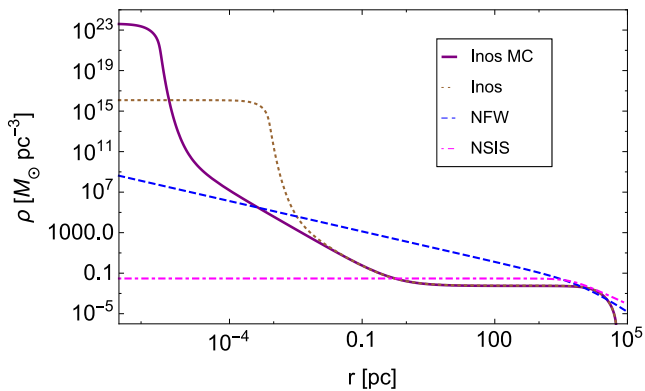}
\caption{Distribution of DM in MW-type galaxies predicted by the RAR model. The solid line refers to the most compact (MC) solution for $m=345$~keV (``inos" MC in the legend). For comparison, we show with the dotted brown line, the solution for $m=48$~keV, which we refer to as ``inos" in the legend. We also show the NFW and NSIS profiles given by the formulas (\ref{a1}) and (\ref{a2}), respectively. The free parameters in these profiles were taken from \cite{2013PASJ...65..118S} and
\cite{2009PASJ...61..227S}, respectively, satisfying the
same (total) rotation curve data as in the RAR case, with the corresponding
considerations of bulge and disk counterparts.} \label{density}
\end{figure}

Therefore, the narrow particle mass range provides several solutions to
satisfy either the rotation curve data in the halo part or both sets of
data, namely including additionally the orbits of the S-cluster stars such
as the S2 star, necessary to establish the compactness of the DMCC. A
comparison between the RAR model, NFW profile and NSIS for MW-like spiral
galaxies is also shown in Fig.~\ref{density}, describing the outstanding
inner structure below parsec scale for the inos profile.
\section{Gravitational lensing properties}\label{sec:3}
As the RAR density profile cannot be given analytically,
we here compute numerically the general GL properties for this model. The
lensing effect causes the image of the source to be displaced, magnified
(or demagnified), and sometimes splitted. Interestingly, these effects can be
quantified and contrasted with observations; however, we do not focus on
any particular lens system but rather on describing the GL properties of
the RAR solutions in nearby MW-type spiral galaxies. To do so, we consider
the particular solutions provided by the particle mass value $m=345$~keV
and $m=48$~keV to account also for the central compact object.
In subsections \ref{sec:mass} and
\ref{sec:magnification} we use the standard lensing formalism which is
based on the assumption that the gravitational field is so weak that the
deflection angles are small. In subsection \ref{sec:stronglensing} we
consider light rays that come so close to the central object that this
approximation is no longer valid; there we have to calculate the deflection
angles with the exact equations from general relativity. We also make a
comparison of our results with the respective ones of the NFW and NSIS
profiles as lensing models, in order to infer significant differences that
might help us, along with rotation curve data, to discriminate between
these DM galactic profiles.
%
Henceforth, we will consider in all computations that the source and lens
position are located at $z_{s}=2$ and $z_{l}=0.3$, respectively, which
correspond to typical separations for both the source and the lens. The
cosmological parameters have been taken from the last results of Planck
\cite{2016A&A...594A..13P}: $H=67.80$ km~s$^{-1}$~Mpc$^{-1}$,
$\Omega_{m}=0.3$ and $\Omega_{\Lambda}=0.7$ to determine the angular
diameter distances of the system. For this cosmology, the scale in the lens plane is $1$~arcsec~$=$~$4.74$~kpc.

\subsection{Surface mass density and convergence}\label{sec:mass}
\begin{figure} \centering \includegraphics[width=\hsize,clip]{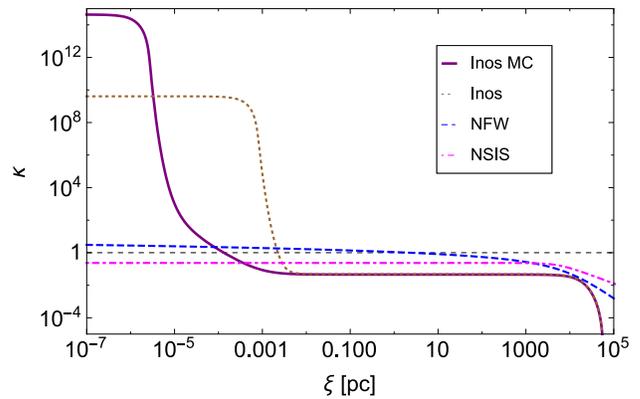}
\caption{Convergence for the NFW, NSIS and RAR density profiles. The
condition of strong lensing is achieved for the RAR profile (inos MC) below
$10^{-4}$~pc while the halo part is characterized by weak lensing effects
as well as for the other profiles.} \label{convergence2} \end{figure}

Considering the lens system as an axially symmetric lens, the planar distribution of matter, i.e. the \textit{projected
surface density}, is obtained by integrating the three-dimensional density
profile $\rho(r)$\footnote{Here the radial coordinate $r$ is related to
cylindrical polar coordinates by $r=\sqrt{\xi^{2}+z^{2}}$.} along the line
of sight \cite{1992grle.book.....S}
\begin{equation} \Sigma(\xi)=2\int_{0}^{\infty} \rho(\xi,z) dz,\label{a5}
\end{equation}
where  $\xi$ is  the \textit{impact parameter} measured from the center of
the lens. For this configuration, the \textit{mean surface density} inside
the radius $\xi$ is
\begin{equation} \bar{\Sigma}(\xi)=\frac{1}{\pi \xi^{2}}\int_{0}^{\xi}2\pi
\xi' \Sigma(\xi')d\xi'.\label{a6} \end{equation}
A useful dimensionless quantity that characterizes the system is the
\textit{convergence}, defined as the ratio of the surface density and the
critical density
\begin{equation} k(\xi)=\frac{\Sigma(\xi)}{\Sigma_{cr}},\label{a11}
\end{equation}
where $\Sigma_{cr}=\frac{c^{2}}{4\pi G} \frac{D_{s}}{D_{l}D_{ls}}$ and $c$
is the speed of light. $D_{s}$, $D_{l}$ and $D_{ls}$ are
the angular diameter distances of the observer to the source, of the
observer to the lens, and of the lens to the source, respectively. Based
on these definitions, we compute the convergence as a function of the
impact parameter $\xi$ for all the above density profiles (see
Fig.~\ref{convergence2}). It is common to give the impact
parameter in units of a reference length $\xi _0$. For the RAR model,
however, $\xi_{0}$ has not yet been identified, in contrast to e.g. the NFW
profile. For this reason, all quantities related to lensing properties are
plotted as a function of the physical impact parameter $\xi$ instead of the
usually used dimensionless radius $\xi/\xi_0$. The influence of the DMCC of
the RAR profile on the convergence can be clearly seen in
Fig.~\ref{convergence2}. If $\kappa>1$, multiple images and Einstein rings
may be formed. It is common to speak of ``strong lensing'' in such
situations. Note, however, that we are still in the regime where the
gravitational field is weak and bending angles are small. From the diagram
we read that for the RAR profile strong lensing effects start to be
notorious at a radius smaller than $10^{-4}$~pc, whereas in the halo part
only weak lensing takes place, for all the density profiles, as expected
\cite{2007MNRAS.374.1427S}.

The density of the RAR profile is, at small distances, several orders of
magnitude higher than that of the NFW and NSIS counterparts. Hence, the
compactness of the DM compact core may eventually lead to the additional
formation of (relativistic) Einstein rings or (relativistic) multiple
images, similarly to the case of a supermassive BH. This possibility will
be analyzed in Sec.~\ref{sec:stronglensing} via a general relativity
treatment beyond the weak-field approximation.

\begin{figure} \centering \includegraphics[width=\hsize,clip]{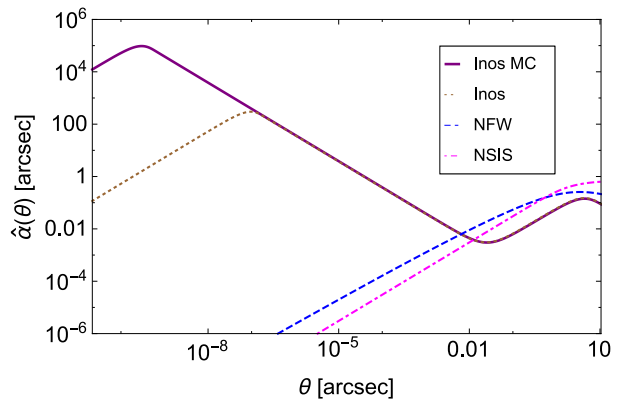}
\caption{Deflection angle for the NFW, NSIS and RAR density profiles. It
can be seen the relation between the deflection angle and the rotation
curve as it was found in \cite{2003PhRvD..68b3516B}. See also
Ref.~\cite{2015MNRAS.451..622R} for the rotation curve behavior.} \label{deflection2} \end{figure}

On the other hand, an interesting quantity that characterizes the system is
the \emph{shear} 
\begin{equation}
\gamma(\xi)=\frac{\bar{\Sigma}({\xi})-\Sigma({\xi})}{\Sigma_{cr}}.
\label{a12} \end{equation}
which determines the distortion of images. For our DM
fermionic configuration, the formation of peaks in the shear (which correspond to the formation of Einstein rings) appears
presumably in zones where the surface density changes abruptly due to the
dominance of DM (in the central and halo part). This feature is similar to that of the deflection angle plotted
in Fig.~\ref{deflection2}.
\subsection{Deflection angle and magnification}\label{sec:magnification}
The deflection angle can be written as
\begin{equation} \hat{\alpha}(\xi)=\frac{4G}{c^{2}}\frac{2\pi
\int_{0}^{\xi}\Sigma(\xi')\xi' d\xi'}{\xi}=\frac{4G
M(\xi)}{c^{2}\xi},\label{a7} \end{equation}
with $M(\xi)$ being the mass enclosed by a circle of radius $\xi$.
For any given density profile $\rho (r)$ we may
(numerically) first calculate $\Sigma ( \xi )$ and then $M ( \xi )$ which,
by (\ref{a7}), gives us the deflection angle $\hat{\alpha}(\xi)$. The
results are shown for the NFW, NSIS and RAR profiles in
Fig.~\ref{deflection2} where the deflection angle is plotted as a function
of the position angle in the sky, $\theta = \xi / D_{l}$. For the RAR
profile it can be observed that the deflection angle becomes larger when a
light ray is close to the DMCC in comparison to one in the halo part,
giving rise to one extra Einstein ring, see below. This is a unique feature
of the GL produced by the RAR profile, since for the other profiles the
maximum deflection angle is obtained in the halo part near the flat part of
the rotation curve. The maximum deflection in the RAR profile case has the
value  $9.49\times 10^{4}$~arcsec ($\hat{\alpha}=0.46$~rad), at radius $0.30$~nano-arcsec ($1.42\times 10^{-6}$~pc) which
is slightly underestimated because in this regime the weak-field
approximation is actually no longer valid. An exact relativistic treatment
will be given in the next section.

In the halo part, all the DM density profiles must fit as a first condition
the rotation curve data.
Moreover, a complementary requirement can be added by considering the light
deflection by the galactic halo. With this extra information at our disposal we
can in principle discriminate between different halo models
\cite{2006MNRAS.372..136F,2010PhRvD..82b4025N} which predict slightly
different deviations of light (of 0.1~arcsec) as can be seen in
Fig.~\ref{deflection2}. On the other hand, starting from 2~mili-arcsec (20~pc)
the RAR profile produces a constantly increasing deflection in logarithmic
scale, toward the central part, reaching the maximum value at $0.30$~nano-arcsec ($1.42\times 10^{-6}$~pc)
due to the DMCC gravitational potential (see Fig.~\ref{deflection2}).

Looking at the halo part, the deflecting angle (in arcsec) for each profile
is computed (according to the best fit parameters) at a distance
$R^{GC}=8.3$ kpc (our distance from the Galactic center) where the circular velocity is near its maximum value
\begin{eqnarray} \hat{\alpha}^{NFW}(R^{GC}) & \approx & 0.26'',\\
\hat{\alpha}^{NSIS}(R^{GC}) &\approx & 0.69'',\label{eq:alphas}\\
\hat{\alpha}^{inos}(R^{GC}) & \approx & 0.15''. \end{eqnarray}
We recall that at such a radius scale the total circular velocity must
fulfill $V_c\approx 220$ km~s$^{-1}$, which implies that lensing data could
serve as a discriminator between dark matter models. Interestingly, it has
been recently inferred the model parameters, i.e. the total (disk+bulge)
stellar mass, the DM halo asymptotic circular velocity, and the core radius
among others, for the spiral galaxy lens SDSS J2141−000 system by using
either strong lensing data, kinematics data, or both combined \cite{2011MNRAS.417.1621D}. This analysis shows that the uncertainty associated with the circular velocity when only the optical emission
and absorption line spectroscopy is considered, is commonly less than
that of strong lensing data (see Table 5 in \cite{2011MNRAS.417.1621D} for
comparison). It implies directly that we can fit very well data from
strong lensing since the RAR profile fits the rotation curves with a good
precision \cite{2016arXiv160607040A}.

\begin{figure} \centering
\includegraphics[width=\hsize,clip]{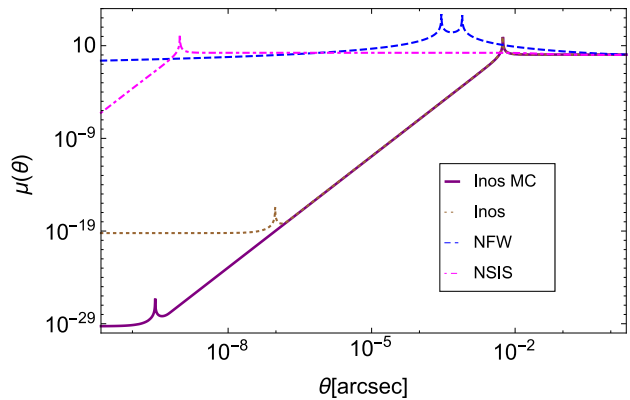} \caption{Magnification
factor for all the profiles listed in the legend and computed by
Eq.~\eqref{a13}.} \label{magnfull} \end{figure}

In addition, the source is also magnified by a factor
\begin{equation}
\mu(\theta)=\frac{1}{(1-k(\theta))^{2}-\gamma(\theta)^{2}}.\label{a13}
\end{equation}
To first order, the \emph{magnification} depends on the convergence only.
Negative values of $\mu$ correspond to inverted images; for large values
of $\theta$, $\mu\rightarrow1$ and the source is weakly affected by the
lensing potential, while for $\theta = \theta_{E}$, the magnification
diverges, corresponding to the formation of an Einstein ring. We calculate
the magnification for all profiles and note the emergence of one extra
Einstein ring due to the DMCC in addition to the halo part. The angular position of such an Einstein ring depends strongly on the compactness of the DMCC subject to the particle mass value between $48$~keV~$\lesssim m c^2 \lesssim 345$~keV. As we can see, the lensing signal is highly demagnified and its effect is indeed comparable to that produced by a SMBH. This remarkable result is plotted in
Fig.~\ref{magnfull}, which clearly shows the effect of the compact DM core
near $10^{-7}$~pc for the more compact solution. This result is also in agreement with the expected demagnified
central image, since the central image flux depends inversely on the square
of the surface density whereby concentrated density profiles should cause
central images to be very faint \cite{2015ApJ...799L..22H,2004Natur.427..613W,2001MNRAS.323..301M}. We stress that Eq.~\eqref{a13} is only valid in the weak field limit; hence, a general description must be used to account for the fully relativistic effects in the more compact solution, i.e. the one derived in the regime of large bending angles. Nevertheless, computing the magnification and the angular position of the additional Einstein ring in the full description of GR\footnote{See for instance section II in \cite{2000PhRvD..62h4003V}, where the exact equations are described for computing the magnification of images in this regime.} provides a value of $\theta_{E}=0.36$~nano-arcsec, which is only 18$\%$ above the one calculated within the weak field limit approximation, i.e. $\theta_{E}=0.30$~nano-arcsec.

It is important to note that we have used in all the calculations the
standard lensing formalism, based on the assumptions that
the gravitational field is weak and that the deflection angle is small. With this, it is possible to
describe the properties of our fermionic DM gravitational lens system,
which is expected to account for such effects very well in the halo 
part\footnote{Such a formalism is still valid for distances far enough away from the DMCC since strong bending effects do not take place.}. However, below $10^{-5}$~pc strong bending must be taken into
account to predict properly the deflection angle due to the DM distribution
in that region. This will significantly affect the lensing
properties of the DMCC. In the next section we calculate these lensing
properties and compare them to those of a black hole.

\subsection{The regime of large  bending angles: Dark
matter central core versus massive black hole}\label{sec:stronglensing}
\begin{figure} \centering
\includegraphics[width=\hsize,clip]{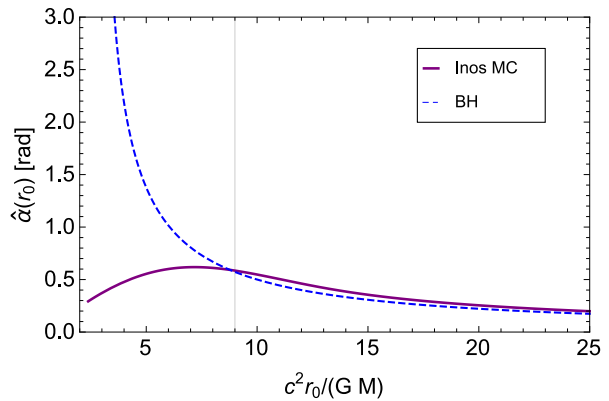} \caption{Comparison
between BH and fermionic DM quantum core of the inos MC configuration. The vertical line indicates the
core radius of the DMCC: $r_{c}\approx 9 G M/c^2$.} \label{fig:strong} \end{figure}

\begin{figure} \centering \includegraphics[width=\hsize,clip]{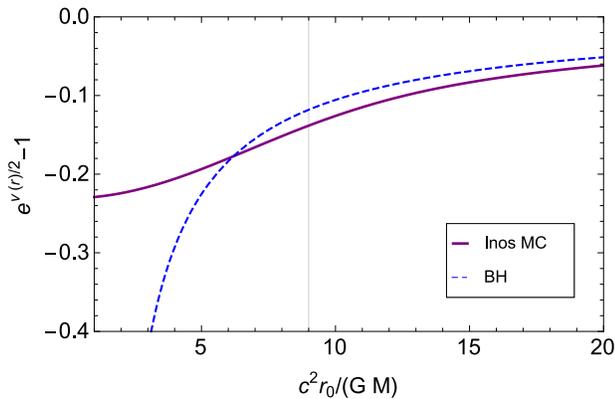}
\caption{Gravitational potential comparison between the central BH and the
fermionic DM configuration. The vertical line indicates the core radius of
the DMCC: $r_{c}=9 G M/c^2$.} \label{fig:potential} \end{figure}

In this part we study lensing due to a fermionic DM core
in a galactic center when a light ray approaches it very closely. For a compact object the bending angle may become large, even exceeding several multiples of $2 \pi$ meaning that the light ray makes several turns around the center. Therefore we have to use the full formalism of general
relativity, beyond the weak-field approximation. Images associated with
light rays that make at least one full turn around the center are often
called ``relativistic images'' although ``higher-order images'' would be a
better nomenclature. We will compare, in the regime where such images occur, the
lensing features of a fermionic DM core with those of a black hole. The
latter have been studied in great detail in the Schwarzschild BH scenario
for Sgr A* \cite{2000PhRvD..62h4003V}, where relativistic images and
relativistic Einstein rings are formed. A natural question that
arises here is whether the DMCC compactness is large enough to account for
the formation of such relativistic images. Hence, we attempt to answer this
question by computing the deflection angle of light rays passing very
closely by the DMCC (and even inside of it since it is treated as
transparent). To do so, we use the formula derived from the static
spherically symmetric metric as a function of the closest light ray
distance of approach $r_{0}$ \cite{1972gcpa.book.....W}
\begin{equation}\label{f1}
\hat{\alpha}(r_{0})=2\int_{r_{0}}^{\infty}\frac{e^{\lambda/2}dr}{\sqrt {(
r^{4}/b^{2}) e^{-\nu}-r^{2}}}-\pi. \end{equation}
where $b$ is the impact parameter:
\begin{equation} b=r_{0}\exp{[-\nu(r_{0}/2)]}. \end{equation}
In the case of the central BH hypothesis (as analyzed in
\cite{2000PhRvD..62h4003V}), the metric outside is described by the
Schwarzschild solution
\begin{equation} e^{\nu(r)}=1-\frac{2 G M}{c^2 r}, \end{equation}
\begin{equation} e^{\lambda(r)}=\left( 1-\frac{2 G M}{c^2 r}\right)^{-1},
\end{equation}
where $M=4.2\times 10^{6} M_{\odot}$ is the BH mass\footnote{For this mass, the Schwarzschild radius, $r_{s}=4.019\times 10^{-7}$~pc.}. Instead, the metric
coefficients for the fermionic model we are interested in here are obtained
by solving the system of equations \eqref{b8}-\eqref{b12} along with
the equations for the density and the pressure, Eqs.~\eqref{b4} and
\eqref{b5}. The result for the deflection angle is plotted in
Fig.~\ref{fig:strong} where we can see that the deflection caused by the DM
central core is small in comparison to that of a black
hole, although considerably beyond the validity of the weak-field formalism
which assumes that $\hat{\alpha}$ may be identified with $\mathrm{tan} \,
\hat{\alpha}$. The maximum deflection, $\hat{\alpha}\approx 0.62$ (distinct
from that obtained by the weak-field approach $\hat{\alpha}=0.46$) is
achieved at $r_{0}\approx 7.18 G M/c^2$ inside the DMCC. Not surprisingly,
a similar feature was observed as in the case of other compact objects such
as fully degenerate fermion stars as well as boson stars
\cite{2000ApJ...537..909B,2000ApJ...535..316D} (see also
Ref.~\cite{Perlick2004} and references therein for a general discussion of
compact objects and their gravitational lensing effects). Interestingly, at distances larger than $r_{c}$, the deflection angles are still appreciably different from the BH ones due to the contribution of the DM distribution leading to a slight difference in its gravitational potential. This is illustrated in Fig.~\ref{fig:potential}.

The deflection angle for the DMCC can be computed approximately by the
Einstein deflection angle provided $r_{0}$ is large, i.e, in the weak field
limit\footnote{It is important to note that this formula is strictly valid
only for the Schwarzschild metric. However, at large radius values the
fermionic solution tends to match the exterior Schwarzschild one.}
\begin{equation} \hat{\alpha}(r_{0})=\frac{4 G M}{c^2
r_{0}}+\mathcal{O}\left(\frac{G^2 M^{2}}{c^4 r_{0}^{2}}\right),\label{f2}
\end{equation}
which at second order implies the value $\hat{\alpha}(r_{0})\approx 0.18$
for $r_{0}=25 G M/c^2$, as can be also seen from Fig.~\ref{fig:strong}. Hence Eq.~\eqref{f2} underestimates by the level of just a few percent the deflection angle compared to that given for the full description Eq.~\eqref{f1} at $r_{0} = 25 GM/c^2$.  Below this radial scale Eq.~\eqref{f2} is no longer valid since it exceeds by more than 10$\%$ the correct relativistic result given by Eq.~\eqref{f1}.

We recall that, for a Schwarzschild BH, relativistic
images are formed due to large bending of light near the photon sphere at
$r=3 G M/c^2$. In this scenario, for the closest distance of approach
$r_{0}=3.21 G M/c^2$ the deflection angle takes the value
$\hat{\alpha}(r_{0})\approx 3\pi/2$ which gives rise to the first
relativistic Einstein ring. This somewhat gives a rough estimate of the
compactness of the lens. By analogy, the compactness of the DMCC can be
obtained as $c^2 r_c/(G M)\approx 9$. However, the light rays can pass
through the DMCC since it does not possess an event horizon, implying a
vanishing deflection angle as the light rays approach the very center of
the configuration. This means that there is no photon sphere, neither
inside nor outside the DMCC.
%
%
This analysis leads to the expected result that the DMCC
does not produce such strong bending effects as a central BH. Therefore, 
the computed deflection of light rays may be used to discriminate the different core compactness, if (highly accurate) observations of the light deflection are available on such short scales. 

Such accurate measurements could be reached in the near future by the Event
Horizon Telescope project for the MW and for M 87.\footnote{http://www.eventhorizontelescope.org}. 
It is also important to note that we only quantified the gravitational
signal effect through the deflection of light, but a more robust study must
be done, i.e. including the motion of stellar or gas components near the MW
center, in order to figure out realistic features that may be discriminant
from observations.

%
\begin{figure} \centering \includegraphics[width=\hsize,clip]{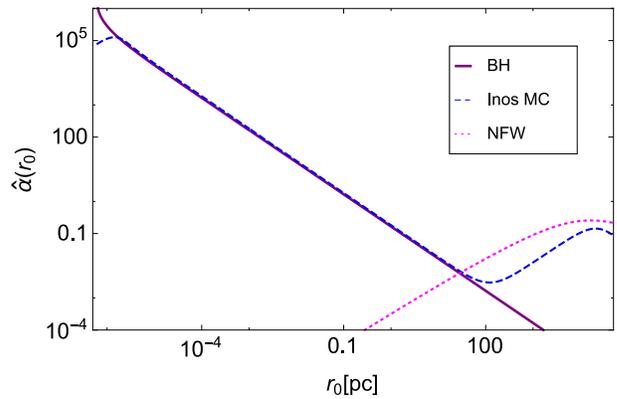}
\caption{Comparison between the BH lensing contribution along the entire
galaxy as well as the most compact solution for the inos profile and the
NFW profile.} \label{fig:strong2}
\end{figure}

Finally, we compute in a general way the deflection angle
both for a BH and for the DMCC along the entire galaxy in order to estimate
the contribution of these compact objects in comparison to the DM halo, see
Fig.~\ref{fig:strong2}. In both cases, we use the exact formula (\ref{f1}).
Apart from a very small region near the center, the two solutions agree
well up to reaching the regime where the contribution from DM halo cannot
be neglected in comparison to the contribution from the BH. Of course, the
fermionic model has to be compared with a \emph{combination} of black hole
and a conventional DM model. 
%
\section{Concluding remarks}\label{sec:4}
In this paper we studied for the first time the gravitational lensing
properties of the fermionic DM distribution (RAR model) in galaxies
\cite{2015MNRAS.451..622R,2016arXiv160607040A}. The RAR model describes
correctly the properties of galactic DM halos of galaxies (including the
flatness of the rotation curves) and, at the same time, it predicts a
denser quantum core towards the center of the distribution. As has been
shown in Ref.~\cite{2016arXiv160607040A}, the compactness of the quantum
core, for a fermion mass in the range $50$~keV~$\lesssim m c^2 \lesssim
345$~keV, is high enough to explain the dynamics of the S-cluster stars,
the closest to the Galactic center. Thus, it represents an alternative
scenario of the central compact object in Sgr A*, traditionally assumed to
be a BH.

We focused on the effect of the DM distribution in the lensing properties
of hypothetical and nearby Milky Way-like spiral galaxies. We first studied
the lensing effects caused by the DM halo region (i.e. neglecting the bulge
and disk contributions to the net lensing). Then we performed the analysis
of the strong bending features near the much denser lens represented by the
DM quantum core at the galaxy center. We computed lensing properties such
as convergence, deflection angle and magnification. We compared and
contrasted the results for the RAR profile with the ones of
phenomenological profiles such as the NFW and the NSIS ones.

For the fermion mass producing the most compact quantum DM core of the RAR
profile (see Fig.~\ref{fig:inosdensity}), $m c^2\approx 345$~keV, which can
explain the Milky Way properties from the center all the way to the halo
\cite{2016arXiv160607040A}, we conclude the following
\begin{enumerate}
\item At distances $r\gtrsim 20$~pc from the center of the galaxy, in the
very inner DM halo regions where the diluted fermionic regime settles, the
effect of the central object on the lensing properties, e.g. the deflection
angle, is negligible (see e.g. Figs.~\ref{deflection2} and
\ref{fig:strong2}).
\item Accurate measurements of the deflection angle in regions
$r\gtrsim$~few kpc where DM starts to dominate, together with rotation
curve data, could help to discriminate between different DM models (see
Fig.~\ref{deflection2}).
%
\item The deflection angle at distances $r\lesssim 20$~pc from the center
of the Galaxy increases in the case of the RAR model while it decreases for
the phenomenological ones (see Figs.~\ref{deflection2} and
\ref{fig:strong2}). The reason for this fundamental difference is the
presence of the compact DM quantum core whose effects start to be
appreciable below those distances (see the next conclusions).
\item In the region of $10^{-6}$~pc~$\lesssim r\lesssim 20$~pc, the lensing
effects of a quantum core and a central BH become a theme of unprecedented
precision (see Fig.~\ref{fig:strong}). The reason for this is that
in this region the difference between the  two cases is
very small (see Fig.~\ref{fig:potential}). This implies that the DM
quantum core can affect lensing features such as the critical curves and
the formation (suppression) of an additional (existing) faint central image
in the same fashion as a central BH.
\item The maximum deflection produced by the DM quantum core occurs at
$\textbf{r} \approx 7GM_{c}/c^{2} \approx 1.4\times 10^{-6}$~pc (see
Fig.~\ref{fig:strong}; $M_c$ is the mass of the DM core). This is produced
with a characteristic demagnified signal as in the case of a central BH
(see Fig.~\ref{magnfull}).
\item The effects of strong lensing (multiple images and
Einstein rings) are important at short distances $\sim 10^{-4}$~pc for the more compact solution, when the condition $\kappa>1$ is achieved (see
Fig.~\ref{convergence2}). 
\item Large differences in the deflection angle produced by a DM central
core and a central BH appear at distances $r\lesssim 10^{-6}$~pc (see
Fig.~\ref{fig:strong}). Inside this region the density of the DM quantum
core is nearly constant (see Fig.~\ref{fig:inosdensity}) and its associated
gravitational potential becomes weaker with respect to the one of a BH with
the same mass (see Fig.~\ref{fig:potential}). The reason is that the DMCC
does not possess an event horizon; i.e. it is regular at the center in
contrast to the central BH, implying a vanishing deflection angle as the
light rays approach the very center of the DMCC.
\item The quantum DM core does not show a photon sphere but it can generate
multiple images and Einstein rings (see Fig.~\ref{magnfull}).
Interestingly, the proposed Event Horizon Telescope uses a Very Long
Baseline Interferometry array of (sub)millimeter
telescopes that could resolve the predicted shadow of the central BH within
the next years with the inclusion of the Atacama Large
Millimeter/submillimeter Array (ALMA). The expected angular resolution is
20--30~$\mu$arcsec \cite{2009astro2010S..68D}, whereas the
predicted angular diameter of the shadow is 54~$\mu$arcsec. If a BH shadow
will not be observed, then it will open a window for alternative scenarios
regarding the nature of the SgrA* central object including the DM quantum
core predicted by the RAR model.
\end{enumerate}

Analogous conclusions apply as well to the RAR profiles obtained for other
fermion masses in the range $50$~keV~$\lesssim m c^2 \lesssim 345$~keV and
for other galaxy types such as dwarf and elliptical galaxies, due to the
universal behavior of the RAR density profiles (see Fig.~2 in
\cite{2016arXiv160607040A}). The latter opens the interesting possibility
to use the lensing data for single galaxies from surveys such as SWELLS
\cite{2011MNRAS.417.1621D,2012MNRAS.423.1073B} and DiskMass
\cite{2013A&A...557A.131M}.

We have considered in this work the gravitational lensing produced by
fermionic DM distributions within the RAR model for isolated galaxies.
Since the lensing is enhanced in clusters of galaxies, the generalization
of the RAR profile in the presence of galaxy interactions deserves to be
explored in future works.

In Ref.~\cite{2016arXiv160607040A} it has been shown that, for a fermion
mass range $50$~keV~$\lesssim m c^2 \lesssim 345$~keV, the RAR profile is
consistent both with the Milky Way data and, when applied to other
galaxies, with observed galaxy correlations such as the $M_{\rm BH}-M_{\rm
DM}$ relation and the constancy of the central surface DM density.
Interestingly, in the case of a fermion mass of $m c^2\approx 50$~keV, the
DM core becomes gravitationally unstable to BH formation when it reaches a
mass of $\approx 2.3\times 10^8~M_\odot$. This led to the hypothesis made
in Ref.~\cite{2016arXiv160607040A} that supermassive BHs ($M\gtrsim
10^8~M_\odot$) hosted at the center of active galaxies could be formed from
a BH seed given by this DM collapse. Such a newly born BH, soon after its
formation, can accrete baryonic and DM from its surroundings. In that case,
the fermionic DM density profile will be affected by the presence of, and
accretion onto, the central BH. We are planning to perform an analysis of
the lensing properties of such a accretion-modified RAR profile, as well as
its feedback on the BH shadow properties (see e.g.
Ref.~\cite{2013A&A...554A..36L} for the case of a NFW density profile), in
a future publication.

\section{Acknowledgments}
L.G.G. is supported by the Erasmus Mundus Joint Doctorate Program by Grant
Number 2013--1471 from  the EACEA  of the European Commission. L.G.G is
grateful for the warm hospitality at ZARM Institute during his mobility
period. C.R.A acknowledges support from ICRANet and CONICET-Argentina.
J.A.R. acknowledges the support by the International Cooperation Program
CAPES-ICRANet financed by CAPES-Brazilian Federal Agency for Support and
Evaluation of Graduate Education--within the Ministry of Education of
Brazil.

%

%

\bibliographystyle{apsrev4-1} \end{document}